\documentclass[journal=jacsat,manuscript=article]{achemso}

\usepackage[version=3]{mhchem} 
\usepackage{siunitx}

\usepackage{graphicx} 
\usepackage{amsmath}
\DeclareMathOperator\arccot{arccot}

\title{Perfect absorption at the ultimate thickness limit in planar films}
\author{Zarko Sakotic}
\email{zarko.sakotic@austin.utexas.edu}
\author{Alexander Ware}
\affiliation[UT Austin]{Electrical and Computer Engineering, University of Texas at Austin, Austin, TX 78758, United States}
\author{Michelle Povinelli}
\affiliation[USC]{Ming Hsieh Department of Electrical and Computer Engineering, University of Southern California, Los Angeles, California 90089, United States}
\author{Daniel Wasserman}
\affiliation[UT Austin]
{Electrical and Computer Engineering, University of Texas at Austin, Austin, TX 78758, United States}
\email{dw@utexas.edu}
\date{July 2023}

\begin{document}

\maketitle

\begin{abstract}
  Reducing device volume is one of the key requirements for advanced nanophotonic technologies, however this demand is often at odds with designing highly absorbing elements which usually require sizeable thicknesses, such as for detector and sensor applications. Here we theoretically explore the thickness limitations of perfectly absorbing resonant systems and show surprisingly low bounds on minimal required thicknesses for total light absorption in thin planar films. We present a framework for understanding, predicting, and engineering topologically protected perfect absorption in a wide range of resonantly absorbing materials. The proposed analytical approach leads to a simple relation between a perfect absorber’s thickness and dielectric function loss, which also serves as a guide for determining the absorption potential of existing and emerging materials at the ultimate thickness limit. The presented results offer new insights into the extremes of light-matter interaction and can facilitate the design of ultra-sensitive light absorbers for detector and sensor systems.
\end{abstract}

\section{Introduction}

Maximizing absorption of electromagnetic waves in thin material layers has been a topic of research for almost a century \cite{woltersdorff1934optischen,dallenbach1938reflection, hadley1947reflection,salisbury1952absorbent, fante1988reflection}. The study of interference in thin films has been given considerable attention since then, as micro- and nano-fabrication techniques have advanced substantially in the past decades, enabling implementation and interrogation of nano-scale structures. Strong absorption of light in thin layers is at the center of a number of applications such as thermal emission engineering \cite{greffet2002coherent}, photodetectors \cite{landy2008perfect,li2014metamaterial}, sensors\cite{liu2010infrared}, solar absorbers \cite{ferry2008plasmonic,pala2009design}, optical coatings\cite{kats2012ultra}, modulators \cite{yao2014electrically} and quantum transduction for single photon detection\cite{akhlaghi2015waveguide, vetter2016cavity, kokkoniemi2020bolometer}. With the advent of metamaterials and metasurfaces, a host of new absorber designs have been introduced  \cite{ra2015thin,yu2019broadband}, expanding the absorbing structures' functionalities and applicability. Considerable efforts in this area have been focused towards engineering perfect absorption (PA) \cite{watts2012metamaterial,thongrattanasiri2012complete}, sometimes referred to as critical coupling\cite{yariv2002critical, piper2014total}, existing at only a single point in a parameter space. These special points are usually observed in one-port systems by measuring reflection zeros\cite{kats2016optical}, as other scattering channels (transmission, diffraction) are suppressed by design, i.e. by using mirrors, operating in the total internal reflection regime, or by tailoring a suitable metamaterial response.

The PA phenomenon was further generalized to multiple-port, open scattering systems\cite{chong2010coherent}, best understood as a monochromatic solution to a boundary wave problem with only incoming waves, coined coherent perfect absorption (CPA) or anti-lasing. The name comes from the fact that, in an ideal linear system, lasing and CPA are time-reversed versions of each other, as the onset of lasing is signified by a self-sustained (source-free) solution of the boundary problem, i.e. only outgoing waves exist with no input or source. This duality is also reflected in their representation as real-frequency poles and zeros of the scattering matrix eigenvalues\cite{baranov2017coherent, krasnok2019anomalies}, which carry notable topological properties as singularities of the scattering response \cite{kravets2013singular,berkhout2019perfect,sakotic2021topological,sakotic2023non,colom2023crossing,Liu2023TopologicalPA}, thereby introducing a topological aspect to PA. With the recent emergence of 2D materials, engineering PA in few-layer and monolayer materials such as transition metal dichalcogenides (TMDCs) became possible \cite{jariwala2016near,epstein2020near,horng2020perfect}, where the appearance of PA has also been connected to its topological origin \cite{ermolaev2022topological,canales2023perfect}. Remarkably, strong optical response in such materials persists even in the monolayer limit \cite{alcaraz2018probing,rivera2019phonon,ma2022reststrahlen}, enabling strong-light matter interactions at the atomic scale. Reaching maximum absorption of light at the ultimate thickness limit enables a path towards the next generation of ultra-sensitive detectors \cite{luhmann2020ultrathin, zhao2023ultrathin,huo2018recent, akinwande2019graphene}, and opens possibilities for studying phenomena intrinsic to quantum information and communication technologies \cite{turunen2022quantum}. Despite the well-established understanding of PA and the recent demonstrations of PA in ultra-thin layers, leveraging an ever-expanding material library, a general material-agnostic framework for resonant perfect absorption at the ultra-thin limit is still lacking. An analytical framework outlining the thickness limitations intrinsic to achieving (near)perfect absorption is necessary for maximizing the potential of absorbing elements in photonic and opto-electronic devices. Furthermore, fundamental studies on maximal light-matter interaction at extreme physical limits is of interest to photonics research at large\cite{yu2010fundamental,miller2016fundamental,molesky2019t,kuang2020maximal,chao2022physical}.

Motivated by this rationale, in this work we propose a new mechanism for understanding, predicting, and engineering perfect resonant absorption at the ultimate ultra-thin limit. We first analytically discuss the emergence of perfect absorption as a topological singularity in a thin, planar resonant layer, and derive the limits of its existence accounting for thickness and material properties. We find a general solution for perfect absorption in thin resonant materials that directly relates the absorbing material thickness to its dielectric function loss, which elucidates the ultimate thickness limits of the perfect absorption phenomenon. These relations do not only predict perfect absorbers of ultra-subwavelength thicknesses, but also reveal that the thickness limit is completely independent of the real part of permittivity of the absorbing material. As a consequence, this finding provides a substantially simplified analytical framework for qualitative assessment of absorption properties at the ultra-thin limit. Our approach applies to all planar, resonantly absorbing materials in the linear regime, including bulk and layered phononic and excitonic materials, as well as metamaterial systems which can be approximated with effective response. We also show that our framework can be extended to high-Q cavities, enabling a  path towards orders of magnitude further reduction in absorber thickness.

\section{Results}

We start our considerations with the analysis of a simple optical problem in three different scenarios - a deeply sub-wavelength absorbing layer of thickness $t$ is: (i) backed by quarter-wave spacer and a perfect electric conductor (PEC); (ii) backed by a perfect magnetic conductor (PMC); (iii) suspended in air, as shown in Fig. \ref{Figure:1}(a i-iii). PEC and PMC here represent perfect mirrors providing reflection coefficient with unity amplitude and phase either $\pi$ or 0, respectively. We suppose the absorbing layer is characterized by a single, narrowband optical transition with no other resonances within the range $\Delta\lambda\approx\lambda_0/5$ - the optical properties of such media are well described by a Lorentz resonance model, effectively used to describe the optical response of phonons in polar dielectrics, excitons in semiconductors, effective metamaterial response, or collective resonances of an array of emitters/absorbers. Specifically, we choose a TO-LO Lorentzian model that is most commonly used for phononic response:

\begin{equation}{\epsilon} =\epsilon_r \epsilon_o =  {\epsilon_\infty}(1+\frac{\omega_{LO}^2-\omega_{TO}^2}{\omega_{TO}^2-\omega^2-{i}\gamma\omega}), \end{equation}

where $\epsilon_\infty$  is the high frequency limit permittivity, $\omega_{LO}$ and $\omega_{TO}$ are the longitudinal and transverse optical phonon frequencies, and $\gamma$=$g$$\omega_{TO}$ is the damping of the oscillator model. The real and imaginary permittivity of such an absorbing material are shown in Fig. \ref{Figure:1}(b). A simple approach in analyzing single- and multi-layer problems is based on using wave impedance, where reflection and transmission can be calculated using a transfer matrix model [suppl. doc.]. Since the layer in question is much thinner than the wavelength, we can describe its impedance with sheet conductivity \cite{li2018two,ma2022reststrahlen}, which further simplifies our analysis. Reflection and transmission coefficients for the suspended layer can thus be written as:

\begin{figure} [t]
\centering
\fbox{\includegraphics[width=16cm]{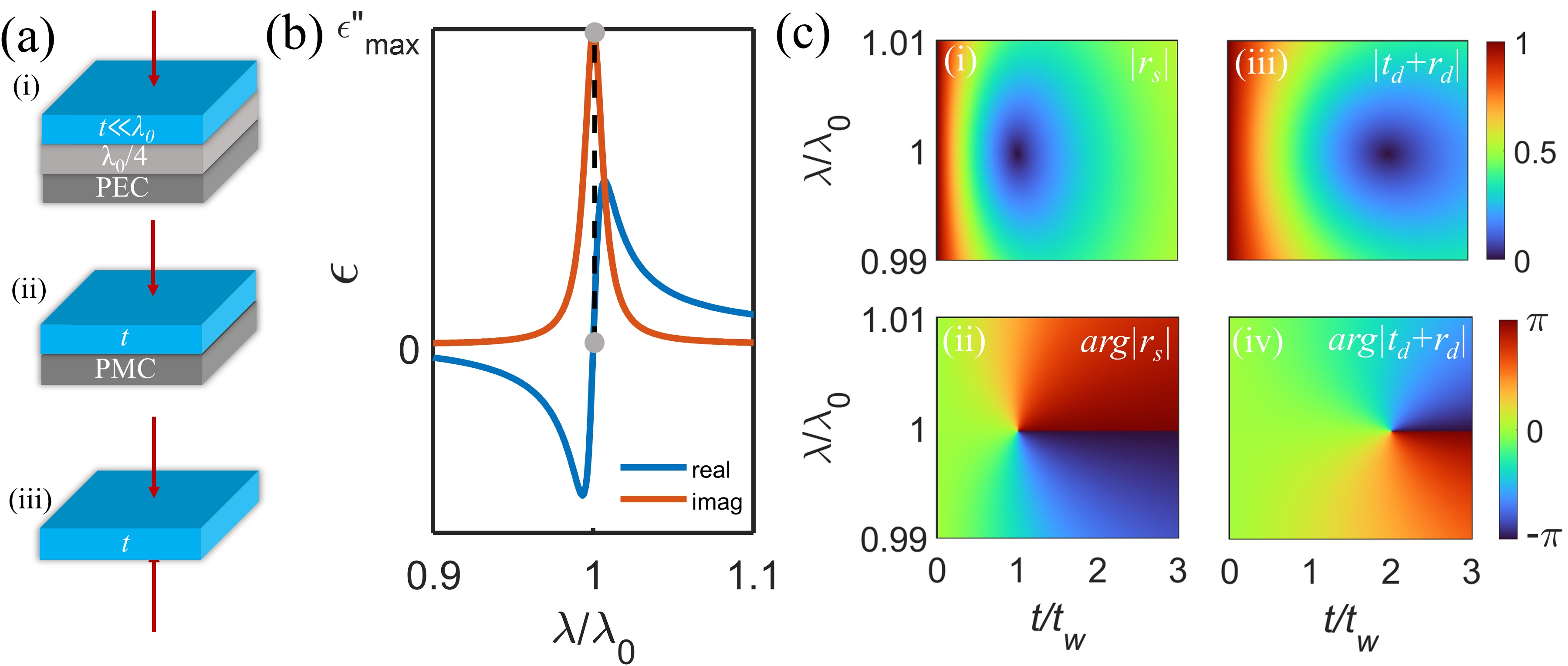}}
\caption{(a) Schematics of the three perfect absorption configurations under consideration. (b) Real and imaginary part of the dielectric function describing the optical response of the representative thin film in the vicinity of a Lorentzian resonance with the peak of the imaginary part of the permittivity highlighted. (c) Reflection coefficient amplitude (i) and phase (ii) for the PMC-backed thin resonant film. $S$-matrix eigenvalue amplitude (iii) and phase (iv) for the freestanding thin resonant film.}
\label{Figure:1}
\end{figure}

\begin{equation}{r_d} = \frac{-{Z_0}/{Z_s}}{2+{Z_0}/{Z_s}}, \end{equation} \begin{equation} 
{t_d} = \frac{2}{2+{Z_0}/{Z_s}} \end{equation}
 
 where ${Z_0}$ is the free space impedance ($\approx$377$\Omega$), \begin{math}Z_s = iZ_0\lambda/((\epsilon_r-1)2\pi{t}) \end{math} is the absorber layer sheet impedance\cite{li2018two}, and \begin{math}\epsilon_r=\epsilon'+i\epsilon'' \end{math} is the complex relative permittivity of the thin layer. For the one-port, PMC-backed configuration the reflection coefficient is simply:
 
\begin{equation}{r_s} = \frac{{Z_s}-{Z_0}}{{Z_s}+{Z_0}}.\end{equation}
 
 Note that, at resonance, the reflection coefficient for the quarter wave spacer backed by PEC is identical to eq. (4), thus for brevity we only discuss the PMC case in the following discussion. We plot the amplitude and phase of the reflection coefficient ${r_s}$ in the spectral vicinity of the absorption resonance, as a function of thickness and wavelength, in Fig. \ref{Figure:1}(c, i-ii). As can be seen from Fig. \ref{Figure:1}(c, i-ii), there is a singular point where reflection goes to zero (topological phase singularity), indicating perfect absorption. This happens exactly at $\lambda_0$, for a specific thickness ${t_w}$. As we further show, this thickness is a version of the previously derived Woltersdorff thickness \cite{woltersdorff1934optischen, pu2012ultrathin}, which from here on we term the \textit{resonant Woltersdorff thickness}, and which has direct and important implications for thickness limitations in perfectly absorbing resonant systems. Originally, the Woltersdorff thickness was derived for suspended or mirror-backed metal films in the low frequency regime where the real and imaginary parts of the refractive index are large and roughly equal \begin{math}n \approx k \gg 1 \end{math}\cite{pu2012ultrathin, luo2014unified, li2014equivalent}, i.e. where the imaginary part of the permittivity dominates. Similar calculations have led to the so-called Berreman thickness\cite{harbecke1985optical} and plasmon thickness \cite{luo2018subwavelength}, for which the maximum of absorption is achieved at or near the plasma frequency, where real part of the permittivity goes through zero (epsilon-near-zero, ENZ), and permittivity is purely (or mostly) imaginary. Surprisingly, however, little attention has been paid to the fact that a similar permittivity scenario happens at the resonance of the material. Namely, at the resonant wavelength $\lambda_0$, the real part of the permittivity crosses zero while the imaginary part is at its maximum, leading to a diverging $\epsilon''/\epsilon'$ ratio - a large value of this ratio has usually been regarded as a requirement for impedance-matched ultra-thin absorbers \cite{luo2014unified,kim2016superabsorbing,luhmann2020ultrathin}. At resonance, the term \begin{math} (\epsilon_r-1)\end{math} is purely imaginary and consequently the surface impedance of the layer is purely real, thus it can be matched to free-space for a specific thickness, i.e. for $t_w$. This results in a direct link between the thickness of the layer $t$ and the imaginary part of the permittivity at its peak, i.e. at the resonance. In the mirror-backed film scenario, reflection zero and hence PA happens when the impedance of the film is matched to free space impedance $Z_s$=$Z_0$, which leads to the expression for the normalized resonant Woltersdorff thickness:

\begin{equation} \label{eq5} \frac{t_w}{\lambda_0} = \frac{1}{2\pi\epsilon''_{max}}\end{equation}
 
 where \begin{math}\epsilon''_{max}=g/[2\pi\epsilon_\infty(\omega_{LO}^2/\omega_{TO}^2-1)]\end{math}, specific for the TO-LO model used. The main result in eq. (5), however, is general for all materials with Lorentzian response models, including isotropic and anisotropic materials with in-plane resonant response, as long as the resonance is strong enough so that real part of permittivity drops below 1 around the resonance, i.e. the material effectively contains a Reststrahlen band. When considering oblique incidence angles, the impedance matching condition can be written for two polarization cases: \begin{math}t_w^{TE}/\lambda=\cos{\theta_0}/(2\pi\cos{\theta_1}(\epsilon_r-1))\end{math} and \begin{math}t_w^{TM}/\lambda=\cos{\theta_1}/(2\pi\cos{\theta_0}(\epsilon_r-1))\end{math}. This dependence implies that, for TE incidence, it is possible to achieve PA for thicknesses even smaller than $t_w$. 
 
\begin{figure} [t]
\centering
\fbox{\includegraphics[width=8cm]{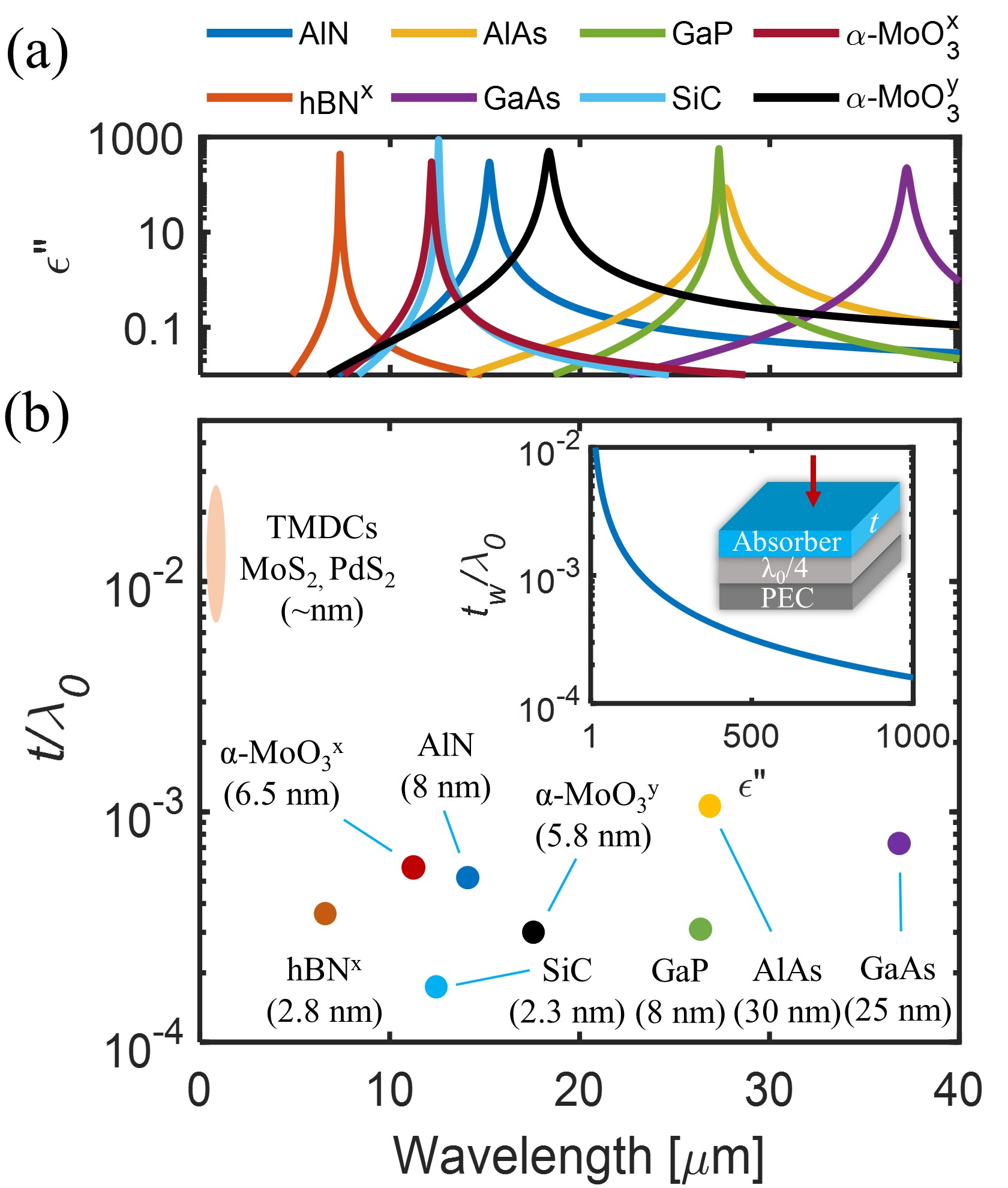}}
\caption{(a) Imaginary permittivity (log scale) for a range of semiconductor materials. (b) Normalized resonant Woltersdorff thickness $t_{w}$ in the PMC configuration with absolute thicknesses in brackets. Inset shows $t_{wn}$ as a function of $\epsilon''_{max}$.}
\label{Figure:2}
\end{figure}

 For the suspended film (in Fig. \ref{Figure:1}(a-ii), which has two input and output scattering channels, the maximum absorption achieved with a single  incident wave is 50\% - this happens when impedance of the layer is half of the free-space impedance $Z_s$=$Z_0$/2\cite{hadley1947reflection, piper2014total, luo2014unified}. Thus, the resonant Woltersdorff thickness is twice the derived value in eq. (5) for the suspended layer. The topological signature of this condition can be seen when plotting the amplitude and phase of one of the scattering matrix eigenvalues \begin{math}s_1 = t_d+r_d\end{math}, as this point corresponds to a CPA or zero of $s_1$, Fig. \ref{Figure:1}(c, iii-iv). As eq. (5) shows, the thickness at which total absorption happens is inversely proportional to the maximum of the imaginary part of the permittivity, which suggests the possibility of total absorption in extremely sub-wavelength layers in a number of different material systems. In Fig. \ref{Figure:2}, we plot the imaginary permittivity and corresponding normalized $t_w$ for a range of semiconductor materials using bulk permittivity from the literature, spanning from the visible to the far-IR regions of the electromagnetic spectrum. Polar dielectrics such as SiC and AlN, layered van der Waals materials such as hBN and $\alpha$-MoO3, and semiconductors such as GaAs and AlAs have strong TO resonances where the imaginary part of the permittivity reaches values of several hundreds, remarkably implying perfect absorption in thicknesses less than $\lambda$/1000. This result does not take into account any potential thickness dependence of the permittivity, which could play a role at such thicknesses; however optical response can remain quite strong (bulk-like), in extremely thin layers in Van der Waals and/or phononic materials, down to nano-scale thicknesses\cite{ma2022reststrahlen,beliaev2021thickness}. This is especially true in monolayer and few-layer TMDCs \cite{ermolaev2022topological,canales2023perfect}, indicating remarkable potential for light-matter interaction at the ultimate ultra-thin limit.

 Here we note that the derived formula in eq. (5) describes PA at normal incidence and only at resonance, where sheet impedance is purely real. However it is also possible to engineer PA at wavelengths around the resonance (and at varying incident angles), where permittivity, and consequently sheet impedance, acquire complex values - this can be done by manipulating the phase of the wave reflected of off the mirror (or the phase of the secondary wave in the free-standing configuration) to induce perfect destructive interference i.e. impedance matching. This could be of particular importance for materials such as TMDCs, where the optical resonance might not be strong enough to induce a Reststrahlen band ($\epsilon{'}$ does not drop to zero) but where strong excitonic transitions nevertheless allow for mono-layer and few-layer perfect absorption. As we discuss further, PA is a topological phase singularity which traverses the parameter space and can be tuned by designing the optical environment of the thin layer, as well as the thickness of absorber itself.  
 
\begin{figure} [t]
\centering
\fbox{\includegraphics[width=16.2cm]{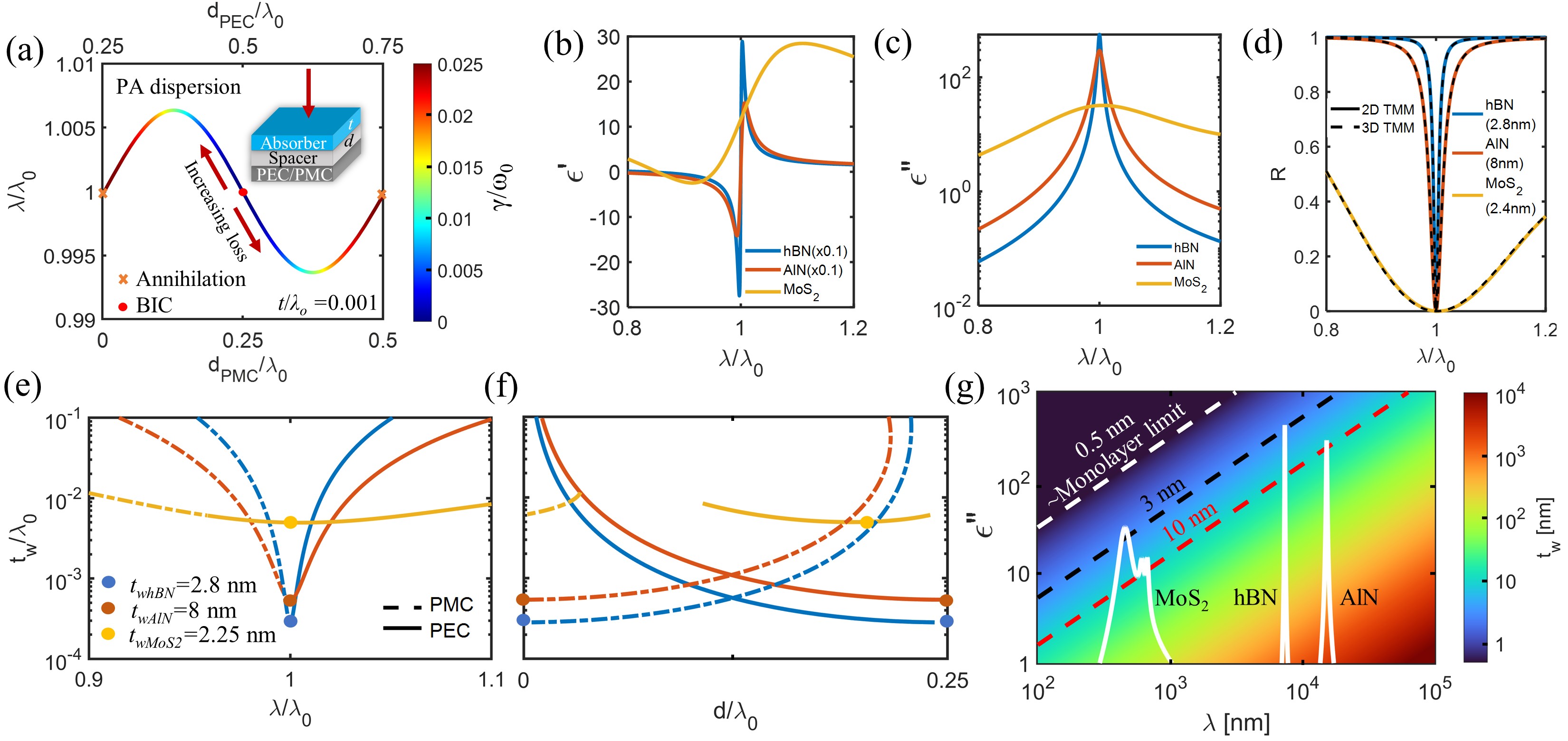}}
\caption{(a) Dispersion of perfect absorption points as a function of $\gamma$ for PEC/PMC-backed cavity. (b)  Real and (c) imaginary permittivities of $MoS_2$, hBN, and AlN from literature. (d) Simulated reflection coefficient for the three materials at the $t_w$ limit in the PEC-backed spacer configuration, with excellent agreement between 2D and 3D TMM models. (e) Dispersion of PA solutions using eqs. (8-10) projected on to the $\lambda-t_w$ and (f) $d-t_w$ parameter space (g) Contour plot of PA thickness from eq. (6) as a function of absolute wavelength and $\epsilon''$, with the dispersion of $\epsilon''$ for the three materials overlaid.}
\label{fig:3}
\end{figure}

  To elucidate this behavior, we now analyze PMC- and PEC-backed lossless (but leaky) dielectric cavities, capped by the resonant absorber ultra-thin layer, Fig. \ref{fig:3}(a, inset), with the same permittivity model as in Fig. \ref{Figure:1}(b). We write the transfer matrix reflection coefficient, and scan the 3-dimensional parameter space consisting of wavelength, dielectric spacer thickness d, and dielectric loss $\gamma$, and numerically locate reflection zeros for fixed absorber thickness $t/\lambda_0=0.001$, Fig. \ref{fig:3} (a). Interestingly, when losses are turned to zero, two PA singularities collapse into a bound state in continuum (BIC) for $d/\lambda_0=0.25$ (PMC case) and $d/\lambda_0=0.5$ (PEC case), as detailed in \cite{sakotic2021topological,sakotic2023non}. This is because in the lossless case, the Lorentzian model response provides PEC-like behavior of the absorber layer at resonance, enabling a dark Fabry-Perot mode between the top and bottom perfect mirrors. When losses are increased from zero, the BIC response inevitably breaks into two topological PA charges\cite{sakotic2021topological}, which traverse the parameter space in a sinusoidal fashion until they annihilate with an oppositely charge PA at $d/\lambda_0=0$ (PMC case) and $d/\lambda_0=0.25$ (PEC case). These two scenarios are identical, and correspond to the well-known Salisbury screen configuration\cite{salisbury1952absorbent,fante1988reflection}. More details on the topological charge creation, dispersion, and annihilation can be found in [suppl. docu.]. This analysis thus reveals that the origin of the topological PA charge is the BIC, while the terminal point is the classic Salisbury screen condition, i.e. for which the resonant Woltersdorff thickness is derived. 

  To further investigate the PA dispersion and its consequences for real absorber materials, we next analyze three illustrative cases in the PMC/PEC-spacer configuration (free-standing configuration discussion can be found in the suppl. doc.). We plot the bulk permittivity of hBN, AlN and $MoS_2$ found in literature, which represent resonantly absorbing materials spanning from visible to beyond the long wave infrared, and with vastly different levels of dielectric loss, Fig. \ref{fig:3}(b,c). For each material, we normalize the wavelength to its resonant wavelength (in-plane TO-phonon resonances for hBN and AlN, and C-exciton resonance for $MoS_2$). Using the same transfer matrix formalism, we find the PA solution for the PMC and PEC cases respectively to be:

\begin{equation}{t_{PMC}} = {\lambda} \frac{i+\tan(2\pi d/\lambda)}{2\pi(\epsilon_r-1)}, \end{equation}
\begin{equation}{t_{PEC}} = {\lambda}\frac{i-\cot(2\pi d/\lambda)}{2\pi(\epsilon_r-1)}. \end{equation}
 
Note that these equations revert back to eq. (5) at resonance for $d$=0 and $d$=0.25$\lambda_0$, respectively.  By forcing $t$ to be real and positive in eqs. (6,7), the PA solutions can be expressed as:
\begin{equation} \label{eq8} {t_{PMC}} = {t_{PEC}} =  \frac{\lambda}{2\pi\epsilon''}, \end{equation}
\begin{equation} \label{eq9}  {d_{PMC}} = \frac{\lambda}{2\pi} \arctan(\frac{\epsilon'-1} {\epsilon''}), \end{equation}
\begin{equation} \label{eq10} {d_{PEC}} = \frac{\lambda}{2\pi} \arccot(\frac{1-\epsilon'} {\epsilon''}). \end{equation}
 
 The dispersion of the solutions projected onto 2D parameter space is shown in Fig. \ref{fig:3}(e,f). Remarkably, the absorber thickness solutions for PA remain, as before, completely independent of the real part of the permittivity, which generalizes the previously derived eq. (5). This implies that for any given wavelength $\lambda$, one can always design a perfect absorber of thickness \begin{math}t(\lambda) = \lambda/2\pi\epsilon''(\lambda)\end{math}, while the real part of the permittivity will contribute to determining the phase matching conditions, i.e. the thickness of the cavity is calculated through eqs. (9,10). 
 
 On one hand, this result is surprising - the PA thickness condition is dependent \textit{only} on the imaginary part of the permittivity, which slightly differs from previous discussions on how large of an $\epsilon''/\epsilon'$ ratio is required for impedance-matched ultra-thin absorbers \cite{kats2012ultra,pu2012ultrathin,kim2016superabsorbing,luhmann2020ultrathin}. On the other hand, this is somewhat expected since the only contribution to absorption comes from $\epsilon''$. This result also implies that the presented framework can be extended to non-resonant material response, i.e., metals below their plasma frequency or conductive 2D films such as MXenes in the low frequency regime, which have recently shown to enable wideband absorption at extremely small thicknesses\cite{luhmann2020ultrathin,zhao2023ultrathin}. 
 
 Further tracing the PA condition in \ref{fig:3}(e,f) through the parameter space leads it to its terminal point, i.e. the resonant Woltersdorff thickness. Interestingly, due to significantly higher losses in the dielectric function of $MoS_2$ than the other two materials, the impedance matching conditions are different and the $t_w$ does not happen at the Salisbury screen condition i.e., for $d=0$ (PMC case) and $d/\lambda_0=0.25$ (PEC case), as indicated by the yellow dot in Fig. \ref{fig:3}(f). This is because at the wavelength of the maximum of the imaginary permittivity, the real part of the permittivity is not equal to 1, thus the thickness $d$ of the cavity will be a specific value dictated by eq. (10).
 For the three materials in question, the normalized $t_w$ differ by orders of magnitude, however their absolute thicknesses happen to all be on the order of several nanometers. To confirm the PA at the $t_w$ limit, we plot the simulated reflection coefficient for the three calculated $t_w$ in three material systems, Fig.\ref{fig:3} (d), where zeros are clearly obtained at or near resonant wavelength, and there is an excellent agreement between the sheet impedance 2D model and the 3D transfer matrix model. The calculated thickness $t_{wMoS_2}$ would roughly equate to 4 atomic layers, which aligns well with the recently reported observation of PA in $MoS_2$\cite{canales2023perfect} (PA observed for 3 layers in the total internal reflection regime for angled TE incidence). Additionally, in Fig. \ref{fig:3}(g) we plot the condition eq. (6) as a function of absolute wavelength and imaginary part of the permittivity, and overlay the imaginary permittivities for the three materials. Based on this graph, one can quantitatively determine the potential for perfect absorption at the ultra-thin limit. For example, the spectral position of the imaginary permittivity peak determines the absolute absorber thickness, eq. (8). Thus, although the $MoS_2$ resonance is significantly weaker than hBN or AlN resonances, the fact that it is located in the visible spectrum enables perfect absorption at thicknesses near the mono-layer limit.

In our discussion so far, we have only considered simple mirror-backed ultrathin absorber configurations. However, there are numerous realizations of enhanced or perfect absorption in more complex, structured systems such as metamaterials \cite{thongrattanasiri2012complete,kim2016superabsorbing} or photonic crystals\cite{piper2014total,horng2020perfect}. Specifically, coupling planar ultrathin absorbing layers to high quality factor modes of periodic structures may lead to PA at thicknesses beyond the limits of our discussion to this point, however a general analytical framework addressing such limits has yet to be formulated. To address this fact and understand how the previously discussed $t_w$ limit relates to high-Q systems, we next analyze a simple extension of our previous model shown in Fig. \ref{fig:4}(a,i). This setup can be considered an ``anti-laser" cavity, due to the similarity to a classic Fabry-Perot laser cavity. By enclosing the previously analyzed absorber-spacer-mirror configuration with a secondary high-reflectivity mirror, we can suppress the radiative loss and thereby enhance the Q factor of the cavity. Such mirrors can in practice be realized with, for example, distributed Bragg reflectors which are common in standard laser cavity designs. As the absorber is positioned in the symmetry point of the cavity i.e., at the maximum of the electric field of a cavity mode, higher Q-factors will result in higher field enhancements, which should, in turn, lead to reduction of $t_w$. To analytically express this prediction, we can simplify the structure and place a PMC at the symmetry point of the cavity \ref{fig:4}(a,ii) (similar to the considerations in Figure 1), and write a two-layer reflection coefficient:

\begin{equation} \label{eq11} {r_t} =  \frac{r_m+r_se^{2ikd}}{1+r_mr_se^{2ikd}}, \end{equation}

where $r_m$ is the top mirror reflection coefficient, \begin{math}r_s = (Z_s-Z_0)/(Z_s+Z_0)\end{math} is the reflection coefficient at the absorber layer, $k$ is the wavevector in the propagation direction in the spacer, and $d$ is the spacer thickness. Since we use the 2D sheet impedance model for the absorber layer, we can treat $r_s$ as a single-layer reflection coefficient (calculated through input impedance of a 2D layer backed by PMC). Equating the $r_t$ to zero at resonance, i.e., setting the numerator in Eq. \ref{eq11} to zero leads to \begin{math}r_s = -r_m\end{math}, which gives the Q-enhanced, reduced resonant Woltersdorff thickness:

\begin{figure} [t]
\centering
\fbox{\includegraphics[width=16cm]{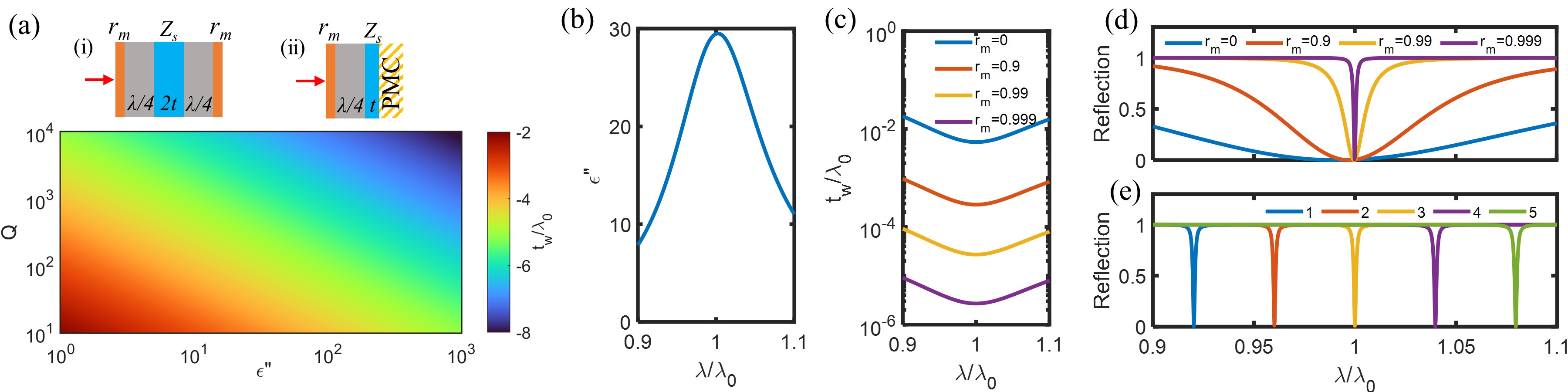}}
\caption{(a) Resonant Woltersdorff thickness as a function of $\epsilon''$ and 
unloaded Q-factor of the cavity. Inset - (i) Sketch of the Q-enhanced, ``anti-laser" cavity with an ultrathin absorber layer; (ii) PMC-backed, simplified version of (i). (b) $\epsilon''$ dispersion of the material under analysis. (c) Dispersion of the normalized $t_w$ for different mirror reflectivities (d) Reflectance dispersion for different mirror reflectivities. (e) Reflectance dispersion with $r_m=0.999$ for different spacer thicknesses $d$ and $t_w$ sampling the spectrum around the resonance.}
\label{fig:4}
\end{figure}

\begin{equation} \label{eq12} {t_{w}^r} = {t_{w}}\frac{1-r_m}{1+r_m} =  \frac{\lambda}{2\pi\epsilon''}\frac{1-r_m}{1+r_m} \approx  \frac{\lambda_0}{4\epsilon''}\frac{1}{Q}, \end{equation}

where \begin{math}Q = \pi r_m/(1-r_m^2)\end{math} is the Q-factor of the uncoupled cavity\cite{siegman1986lasers} (note that setting the denominator to zero in Eq. \ref{eq11} would lead to a reflection pole, i.e., lasing threshold). The inverse relationship of $t_w$ to $\epsilon''$ and $Q$ is 
plotted in Fig. \ref{fig:4}(a). To better illustrate the implications of this result, we further analyze the same system with a realistic absorber material characterized by the $\epsilon''$ shown in Fig. \ref{fig:4}(b), which is similar to the previously analyzed $MoS_2$ in terms of dielectric function loss around the C-exciton. Using Eq. \ref{eq12}, we plot $t_w^r$ and reflectance as a function of wavelength for four different mirror reflection coefficients $r_m$, Fig. \ref{fig:4}(c,d). Remarkably, PA is achieved for orders of magnitude thinner absorbing layers compared to the original structure without the top mirror ($r_m=0$). Similar to considerations in discussion around Figure \ref{fig:3} and Eq. \ref{eq8}, Eq. \ref{eq12} can be generalized to any wavelength around the resonance, while the spacer thickness $d$ needs to be adjusted according to \begin{math} d = \frac{\lambda}{2\pi} \arctan(\epsilon''/(\epsilon'-1)) \end{math}. In principle, one can design arbitrarily thin and narrowband perfect absorbers at any neighboring wavelength around $\lambda_0$, given the availability of an arbitrarily high Q factor and assuming linear material response. As an example of this design possibility, in Fig. \ref{fig:4} (e) we show PA at five different wavelengths (different spacer thicknesses) for $r_m=0.999$ ($Q\approx3*10^4$), where $t_w/\lambda$ is on the order of $10^{-6}$. It is interesting to point out the duality of $t_w$ and lasing threshold calculations - namely, in the linear regime, increasing the Q-factor of the laser cavity decreases the gain threshold, i.e. gain at which lasing starts (first real-frequency pole). Analogously, higher Q-factors decrease the perfect absorber thickness $t_w$, i.e. the anti-lasing threshold (first real-frequency zero). 

\section{Discussion and conclusion}

The presented results imply that $t_w$ is an \textit{intrinsic} thickness limit for perfect resonant absorption in ultra-thin planar films. It represents a material-absorption based thickness bound for extended, non-periodic films, and although derived in three different configurations - freestanding film, Salisbury screen, and high-Q Fabry-Perot cavity - the resonant Woltersdorff thickness $t_w$ in Eqs. \ref{eq5}, \ref{eq8}, and \ref{eq12} has the same underlying material loss dependence. It is important to note that our perfect absorption bound relates the thickness of a single absorbing layer to its loss function around resonance, when the layer is placed in non-absorbing (or weakly-absorbing) optical environments, and without taking into account the total stack thickness or bandwidth. This differentiates our discussion from the Rozanov limit\cite{rozanov2000ultimate}, which relates the strong absorption bandwidth to total stack thickness. Our bounds also differ from recent fundamental studies \cite{kuang2020maximal,venkataram2020fundamental, molesky2019t} which do not consider mirror-backed configurations. For example, our bound in the Salisbury screen configuration predicts minimal PA thicknesses  orders of magnitude smaller than previously derived bound for single-layer, patterned perfect absorbers\cite{kuang2020maximal}, but falls within the general absorption bound derived in \cite{miller2016fundamental}. Our predictions also align well with recent observations in TMDCs \cite{epstein2020near,horng2020perfect,ermolaev2022topological,canales2023perfect} and layered Van der Walls materials\cite{ma2022reststrahlen}, indicating that the analytical approach used here is well-suited for the ultra-thin limit. Although our configuration adds significant thickness to the total structure due to the presence of a spacer, the absorption process is exclusively confined to the ultra-thin layer. Our framework is thus particularly useful for detector design, whose performance strongly benefits from minimizing the absorbing layer thickness itself.  

As long as the use of the 2D/sheet-impedance model is justified and the permittivity model accurately describes the material response, $t_w$ is independent of $\epsilon'$ and inversely proportional to $\epsilon''$. We find that Eqs. \ref{eq5}, \ref{eq8}, and \ref{eq12} accurately hold for $t$/$\lambda<0.05$, i.e. in regions where $\epsilon''>3$. More details on the model accuracy can be found in [suppl. doc]. These considerations lead to very simple analytical relations, ideally suited for qualitatively assessing absorption properties at the ultra-thin limit. Specifically, our approach can be useful when analyzing the absorption potential of emerging novel 2D and phononic materials, as well as providing a new outlook on existing, established materials at the ultra-thin limit.

To conclude, our findings provide new insights on fundamental thickness limitations for maximized light absorption in planar layers. The presented analytical framework shows a simple and direct relation between absorber thickness and material loss, giving a lower bound on achieving perfect absorption. These findings complement fundamental studies of light-matter interaction at the ultimate thickness limits, and provide practical guidance for the design of ultrathin absorbing elements for sensors and detectors. 

\section{Acknowledgement}
{The authors would like to acknowledge support from the Defense Advanced Research Projects Agency (DARPA) under the Optomechanical Thermal Imaging (OpTIm) program (HR00112320022). 
ZS, AW, and DW acknowledge support from Sandia National Labs under the Academic Alliance program. AW also acknowledges support from the DoD NDSEG Fellowship program. This work was performed, in part, at the Center for Integrated Nanotechnologies, an Office of Science User Facility operated for the U.S. Department of Energy (DOE) Office of Science by Los Alamos National Laboratory (Contract 89233218CNA000001) and Sandia National Laboratories (Contract DE-NA-0003525). Sandia National Laboratories is a multimission laboratory managed and operated by National Technology Engineering Solutions of Sandia, LLC, a wholly owned subsidiary of Honeywell International, Inc., for the U.S. DOE’s National Nuclear Security Administration under Contract No. DE-NA-0003525.}

\bibliography{ZarkoUltrathin}

\providecommand{\latin}[1]{#1}
\makeatletter
\providecommand{\doi}
  {\begingroup\let\do\@makeother\dospecials
  \catcode`\{=1 \catcode`\}=2 \doi@aux}
\providecommand{\doi@aux}[1]{\endgroup\texttt{#1}}
\makeatother
\providecommand*\mcitethebibliography{\thebibliography}
\csname @ifundefined\endcsname{endmcitethebibliography}  {\let\endmcitethebibliography\endthebibliography}{}
\begin{mcitethebibliography}{62}
\providecommand*\natexlab[1]{#1}
\providecommand*\mciteSetBstSublistMode[1]{}
\providecommand*\mciteSetBstMaxWidthForm[2]{}
\providecommand*\mciteBstWouldAddEndPuncttrue
  {\def\EndOfBibitem{\unskip.}}
\providecommand*\mciteBstWouldAddEndPunctfalse
  {\let\EndOfBibitem\relax}
\providecommand*\mciteSetBstMidEndSepPunct[3]{}
\providecommand*\mciteSetBstSublistLabelBeginEnd[3]{}
\providecommand*\EndOfBibitem{}
\mciteSetBstSublistMode{f}
\mciteSetBstMaxWidthForm{subitem}{(\alph{mcitesubitemcount})}
\mciteSetBstSublistLabelBeginEnd
  {\mcitemaxwidthsubitemform\space}
  {\relax}
  {\relax}

\bibitem[Woltersdorff(1934)]{woltersdorff1934optischen}
Woltersdorff,~W. Uber die optischen Konstanten d{\"u}nner Metallschichten im langwelligen Ultrarot. \emph{Zeitschrift f{\"u}r Physik} \textbf{1934}, \emph{91}, 230--252\relax
\mciteBstWouldAddEndPuncttrue
\mciteSetBstMidEndSepPunct{\mcitedefaultmidpunct}
{\mcitedefaultendpunct}{\mcitedefaultseppunct}\relax
\EndOfBibitem
\bibitem[Dallenbach and Kleinsteuber(1938)Dallenbach, and Kleinsteuber]{dallenbach1938reflection}
Dallenbach,~W.; Kleinsteuber,~W. Reflection and absorption of decimeter-waves by plane dielectric layers. \emph{Hochfreq. u Elektroak} \textbf{1938}, \emph{51}, 152--156\relax
\mciteBstWouldAddEndPuncttrue
\mciteSetBstMidEndSepPunct{\mcitedefaultmidpunct}
{\mcitedefaultendpunct}{\mcitedefaultseppunct}\relax
\EndOfBibitem
\bibitem[Hadley and Dennison(1947)Hadley, and Dennison]{hadley1947reflection}
Hadley,~L.~N.; Dennison,~D. Reflection and transmission interference filters part I. theory. \emph{JOSA} \textbf{1947}, \emph{37}, 451--465\relax
\mciteBstWouldAddEndPuncttrue
\mciteSetBstMidEndSepPunct{\mcitedefaultmidpunct}
{\mcitedefaultendpunct}{\mcitedefaultseppunct}\relax
\EndOfBibitem
\bibitem[Salisbury(1952)]{salisbury1952absorbent}
Salisbury,~W.~W. Absorbent body for electromagnetic waves. 1952; US Patent 2,599,944\relax
\mciteBstWouldAddEndPuncttrue
\mciteSetBstMidEndSepPunct{\mcitedefaultmidpunct}
{\mcitedefaultendpunct}{\mcitedefaultseppunct}\relax
\EndOfBibitem
\bibitem[Fante and McCormack(1988)Fante, and McCormack]{fante1988reflection}
Fante,~R.~L.; McCormack,~M.~T. Reflection properties of the Salisbury screen. \emph{IEEE transactions on antennas and propagation} \textbf{1988}, \emph{36}, 1443--1454\relax
\mciteBstWouldAddEndPuncttrue
\mciteSetBstMidEndSepPunct{\mcitedefaultmidpunct}
{\mcitedefaultendpunct}{\mcitedefaultseppunct}\relax
\EndOfBibitem
\bibitem[Greffet \latin{et~al.}(2002)Greffet, Carminati, Joulain, Mulet, Mainguy, and Chen]{greffet2002coherent}
Greffet,~J.-J.; Carminati,~R.; Joulain,~K.; Mulet,~J.-P.; Mainguy,~S.; Chen,~Y. Coherent emission of light by thermal sources. \emph{Nature} \textbf{2002}, \emph{416}, 61--64\relax
\mciteBstWouldAddEndPuncttrue
\mciteSetBstMidEndSepPunct{\mcitedefaultmidpunct}
{\mcitedefaultendpunct}{\mcitedefaultseppunct}\relax
\EndOfBibitem
\bibitem[Landy \latin{et~al.}(2008)Landy, Sajuyigbe, Mock, Smith, and Padilla]{landy2008perfect}
Landy,~N.~I.; Sajuyigbe,~S.; Mock,~J.~J.; Smith,~D.~R.; Padilla,~W.~J. Perfect metamaterial absorber. \emph{Physical review letters} \textbf{2008}, \emph{100}, 207402\relax
\mciteBstWouldAddEndPuncttrue
\mciteSetBstMidEndSepPunct{\mcitedefaultmidpunct}
{\mcitedefaultendpunct}{\mcitedefaultseppunct}\relax
\EndOfBibitem
\bibitem[Li and Valentine(2014)Li, and Valentine]{li2014metamaterial}
Li,~W.; Valentine,~J. Metamaterial perfect absorber based hot electron photodetection. \emph{Nano letters} \textbf{2014}, \emph{14}, 3510--3514\relax
\mciteBstWouldAddEndPuncttrue
\mciteSetBstMidEndSepPunct{\mcitedefaultmidpunct}
{\mcitedefaultendpunct}{\mcitedefaultseppunct}\relax
\EndOfBibitem
\bibitem[Liu \latin{et~al.}(2010)Liu, Mesch, Weiss, Hentschel, and Giessen]{liu2010infrared}
Liu,~N.; Mesch,~M.; Weiss,~T.; Hentschel,~M.; Giessen,~H. Infrared perfect absorber and its application as plasmonic sensor. \emph{Nano letters} \textbf{2010}, \emph{10}, 2342--2348\relax
\mciteBstWouldAddEndPuncttrue
\mciteSetBstMidEndSepPunct{\mcitedefaultmidpunct}
{\mcitedefaultendpunct}{\mcitedefaultseppunct}\relax
\EndOfBibitem
\bibitem[Ferry \latin{et~al.}(2008)Ferry, Sweatlock, Pacifici, and Atwater]{ferry2008plasmonic}
Ferry,~V.~E.; Sweatlock,~L.~A.; Pacifici,~D.; Atwater,~H.~A. Plasmonic nanostructure design for efficient light coupling into solar cells. \emph{Nano letters} \textbf{2008}, \emph{8}, 4391--4397\relax
\mciteBstWouldAddEndPuncttrue
\mciteSetBstMidEndSepPunct{\mcitedefaultmidpunct}
{\mcitedefaultendpunct}{\mcitedefaultseppunct}\relax
\EndOfBibitem
\bibitem[Pala \latin{et~al.}(2009)Pala, White, Barnard, Liu, and Brongersma]{pala2009design}
Pala,~R.~A.; White,~J.; Barnard,~E.; Liu,~J.; Brongersma,~M.~L. Design of plasmonic thin-film solar cells with broadband absorption enhancements. \emph{Advanced materials} \textbf{2009}, \emph{21}, 3504--3509\relax
\mciteBstWouldAddEndPuncttrue
\mciteSetBstMidEndSepPunct{\mcitedefaultmidpunct}
{\mcitedefaultendpunct}{\mcitedefaultseppunct}\relax
\EndOfBibitem
\bibitem[Kats \latin{et~al.}(2012)Kats, Sharma, Lin, Genevet, Blanchard, Yang, Qazilbash, Basov, Ramanathan, and Capasso]{kats2012ultra}
Kats,~M.~A.; Sharma,~D.; Lin,~J.; Genevet,~P.; Blanchard,~R.; Yang,~Z.; Qazilbash,~M.~M.; Basov,~D.; Ramanathan,~S.; Capasso,~F. Ultra-thin perfect absorber employing a tunable phase change material. \emph{Applied Physics Letters} \textbf{2012}, \emph{101}, 221101\relax
\mciteBstWouldAddEndPuncttrue
\mciteSetBstMidEndSepPunct{\mcitedefaultmidpunct}
{\mcitedefaultendpunct}{\mcitedefaultseppunct}\relax
\EndOfBibitem
\bibitem[Yao \latin{et~al.}(2014)Yao, Shankar, Kats, Song, Kong, Loncar, and Capasso]{yao2014electrically}
Yao,~Y.; Shankar,~R.; Kats,~M.~A.; Song,~Y.; Kong,~J.; Loncar,~M.; Capasso,~F. Electrically tunable metasurface perfect absorbers for ultrathin mid-infrared optical modulators. \emph{Nano letters} \textbf{2014}, \emph{14}, 6526--6532\relax
\mciteBstWouldAddEndPuncttrue
\mciteSetBstMidEndSepPunct{\mcitedefaultmidpunct}
{\mcitedefaultendpunct}{\mcitedefaultseppunct}\relax
\EndOfBibitem
\bibitem[Akhlaghi \latin{et~al.}(2015)Akhlaghi, Schelew, and Young]{akhlaghi2015waveguide}
Akhlaghi,~M.~K.; Schelew,~E.; Young,~J.~F. Waveguide integrated superconducting single-photon detectors implemented as near-perfect absorbers of coherent radiation. \emph{Nature communications} \textbf{2015}, \emph{6}, 8233\relax
\mciteBstWouldAddEndPuncttrue
\mciteSetBstMidEndSepPunct{\mcitedefaultmidpunct}
{\mcitedefaultendpunct}{\mcitedefaultseppunct}\relax
\EndOfBibitem
\bibitem[Vetter \latin{et~al.}(2016)Vetter, Ferrari, Rath, Alaee, Kahl, Kovalyuk, Diewald, Goltsman, Korneev, Rockstuhl, \latin{et~al.} others]{vetter2016cavity}
Vetter,~A.; Ferrari,~S.; Rath,~P.; Alaee,~R.; Kahl,~O.; Kovalyuk,~V.; Diewald,~S.; Goltsman,~G.~N.; Korneev,~A.; Rockstuhl,~C.; others Cavity-enhanced and ultrafast superconducting single-photon detectors. \emph{Nano letters} \textbf{2016}, \emph{16}, 7085--7092\relax
\mciteBstWouldAddEndPuncttrue
\mciteSetBstMidEndSepPunct{\mcitedefaultmidpunct}
{\mcitedefaultendpunct}{\mcitedefaultseppunct}\relax
\EndOfBibitem
\bibitem[Kokkoniemi \latin{et~al.}(2020)Kokkoniemi, Girard, Hazra, Laitinen, Govenius, Lake, Sallinen, Vesterinen, Partanen, Tan, \latin{et~al.} others]{kokkoniemi2020bolometer}
Kokkoniemi,~R.; Girard,~J.-P.; Hazra,~D.; Laitinen,~A.; Govenius,~J.; Lake,~R.; Sallinen,~I.; Vesterinen,~V.; Partanen,~M.; Tan,~J.; others Bolometer operating at the threshold for circuit quantum electrodynamics. \emph{Nature} \textbf{2020}, \emph{586}, 47--51\relax
\mciteBstWouldAddEndPuncttrue
\mciteSetBstMidEndSepPunct{\mcitedefaultmidpunct}
{\mcitedefaultendpunct}{\mcitedefaultseppunct}\relax
\EndOfBibitem
\bibitem[Ra’di \latin{et~al.}(2015)Ra’di, Simovski, and Tretyakov]{ra2015thin}
Ra’di,~Y.; Simovski,~C.~R.; Tretyakov,~S.~A. Thin perfect absorbers for electromagnetic waves: theory, design, and realizations. \emph{Physical Review Applied} \textbf{2015}, \emph{3}, 037001\relax
\mciteBstWouldAddEndPuncttrue
\mciteSetBstMidEndSepPunct{\mcitedefaultmidpunct}
{\mcitedefaultendpunct}{\mcitedefaultseppunct}\relax
\EndOfBibitem
\bibitem[Yu \latin{et~al.}(2019)Yu, Besteiro, Huang, Wu, Fu, Tan, Jagadish, Wiederrecht, Govorov, and Wang]{yu2019broadband}
Yu,~P.; Besteiro,~L.~V.; Huang,~Y.; Wu,~J.; Fu,~L.; Tan,~H.~H.; Jagadish,~C.; Wiederrecht,~G.~P.; Govorov,~A.~O.; Wang,~Z. Broadband metamaterial absorbers. \emph{Advanced Optical Materials} \textbf{2019}, \emph{7}, 1800995\relax
\mciteBstWouldAddEndPuncttrue
\mciteSetBstMidEndSepPunct{\mcitedefaultmidpunct}
{\mcitedefaultendpunct}{\mcitedefaultseppunct}\relax
\EndOfBibitem
\bibitem[Watts \latin{et~al.}(2012)Watts, Liu, and Padilla]{watts2012metamaterial}
Watts,~C.~M.; Liu,~X.; Padilla,~W.~J. Metamaterial electromagnetic wave absorbers. \emph{Advanced materials} \textbf{2012}, \emph{24}, OP98--OP120\relax
\mciteBstWouldAddEndPuncttrue
\mciteSetBstMidEndSepPunct{\mcitedefaultmidpunct}
{\mcitedefaultendpunct}{\mcitedefaultseppunct}\relax
\EndOfBibitem
\bibitem[Thongrattanasiri \latin{et~al.}(2012)Thongrattanasiri, Koppens, and De~Abajo]{thongrattanasiri2012complete}
Thongrattanasiri,~S.; Koppens,~F.~H.; De~Abajo,~F. J.~G. Complete optical absorption in periodically patterned graphene. \emph{Physical review letters} \textbf{2012}, \emph{108}, 047401\relax
\mciteBstWouldAddEndPuncttrue
\mciteSetBstMidEndSepPunct{\mcitedefaultmidpunct}
{\mcitedefaultendpunct}{\mcitedefaultseppunct}\relax
\EndOfBibitem
\bibitem[Yariv(2002)]{yariv2002critical}
Yariv,~A. Critical coupling and its control in optical waveguide-ring resonator systems. \emph{IEEE Photonics Technology Letters} \textbf{2002}, \emph{14}, 483--485\relax
\mciteBstWouldAddEndPuncttrue
\mciteSetBstMidEndSepPunct{\mcitedefaultmidpunct}
{\mcitedefaultendpunct}{\mcitedefaultseppunct}\relax
\EndOfBibitem
\bibitem[Piper and Fan(2014)Piper, and Fan]{piper2014total}
Piper,~J.~R.; Fan,~S. Total absorption in a graphene monolayer in the optical regime by critical coupling with a photonic crystal guided resonance. \emph{Acs Photonics} \textbf{2014}, \emph{1}, 347--353\relax
\mciteBstWouldAddEndPuncttrue
\mciteSetBstMidEndSepPunct{\mcitedefaultmidpunct}
{\mcitedefaultendpunct}{\mcitedefaultseppunct}\relax
\EndOfBibitem
\bibitem[Kats and Capasso(2016)Kats, and Capasso]{kats2016optical}
Kats,~M.~A.; Capasso,~F. Optical absorbers based on strong interference in ultra-thin films. \emph{Laser \& Photonics Reviews} \textbf{2016}, \emph{10}, 735--749\relax
\mciteBstWouldAddEndPuncttrue
\mciteSetBstMidEndSepPunct{\mcitedefaultmidpunct}
{\mcitedefaultendpunct}{\mcitedefaultseppunct}\relax
\EndOfBibitem
\bibitem[Chong \latin{et~al.}(2010)Chong, Ge, Cao, and Stone]{chong2010coherent}
Chong,~Y.; Ge,~L.; Cao,~H.; Stone,~A.~D. Coherent perfect absorbers: time-reversed lasers. \emph{Physical review letters} \textbf{2010}, \emph{105}, 053901\relax
\mciteBstWouldAddEndPuncttrue
\mciteSetBstMidEndSepPunct{\mcitedefaultmidpunct}
{\mcitedefaultendpunct}{\mcitedefaultseppunct}\relax
\EndOfBibitem
\bibitem[Baranov \latin{et~al.}(2017)Baranov, Krasnok, Shegai, Al{\`u}, and Chong]{baranov2017coherent}
Baranov,~D.~G.; Krasnok,~A.; Shegai,~T.; Al{\`u},~A.; Chong,~Y. Coherent perfect absorbers: linear control of light with light. \emph{Nature Reviews Materials} \textbf{2017}, \emph{2}, 1--14\relax
\mciteBstWouldAddEndPuncttrue
\mciteSetBstMidEndSepPunct{\mcitedefaultmidpunct}
{\mcitedefaultendpunct}{\mcitedefaultseppunct}\relax
\EndOfBibitem
\bibitem[Krasnok \latin{et~al.}(2019)Krasnok, Baranov, Li, Miri, Monticone, and Al{\'u}]{krasnok2019anomalies}
Krasnok,~A.; Baranov,~D.; Li,~H.; Miri,~M.-A.; Monticone,~F.; Al{\'u},~A. Anomalies in light scattering. \emph{Advances in Optics and Photonics} \textbf{2019}, \emph{11}, 892--951\relax
\mciteBstWouldAddEndPuncttrue
\mciteSetBstMidEndSepPunct{\mcitedefaultmidpunct}
{\mcitedefaultendpunct}{\mcitedefaultseppunct}\relax
\EndOfBibitem
\bibitem[Kravets \latin{et~al.}(2013)Kravets, Schedin, Jalil, Britnell, Gorbachev, Ansell, Thackray, Novoselov, Geim, Kabashin, \latin{et~al.} others]{kravets2013singular}
Kravets,~V.; Schedin,~F.; Jalil,~R.; Britnell,~L.; Gorbachev,~R.; Ansell,~D.; Thackray,~B.; Novoselov,~K.; Geim,~A.; Kabashin,~A.~V.; others Singular phase nano-optics in plasmonic metamaterials for label-free single-molecule detection. \emph{Nature materials} \textbf{2013}, \emph{12}, 304--309\relax
\mciteBstWouldAddEndPuncttrue
\mciteSetBstMidEndSepPunct{\mcitedefaultmidpunct}
{\mcitedefaultendpunct}{\mcitedefaultseppunct}\relax
\EndOfBibitem
\bibitem[Berkhout and Koenderink(2019)Berkhout, and Koenderink]{berkhout2019perfect}
Berkhout,~A.; Koenderink,~A.~F. Perfect absorption and phase singularities in plasmon antenna array etalons. \emph{ACS Photonics} \textbf{2019}, \emph{6}, 2917--2925\relax
\mciteBstWouldAddEndPuncttrue
\mciteSetBstMidEndSepPunct{\mcitedefaultmidpunct}
{\mcitedefaultendpunct}{\mcitedefaultseppunct}\relax
\EndOfBibitem
\bibitem[Sakotic \latin{et~al.}(2021)Sakotic, Krasnok, Al{\'u}, and Jankovic]{sakotic2021topological}
Sakotic,~Z.; Krasnok,~A.; Al{\'u},~A.; Jankovic,~N. Topological scattering singularities and embedded eigenstates for polarization control and sensing applications. \emph{Photonics Research} \textbf{2021}, \emph{9}, 1310--1323\relax
\mciteBstWouldAddEndPuncttrue
\mciteSetBstMidEndSepPunct{\mcitedefaultmidpunct}
{\mcitedefaultendpunct}{\mcitedefaultseppunct}\relax
\EndOfBibitem
\bibitem[Sakotic \latin{et~al.}(2023)Sakotic, Stankovic, Bengin, Krasnok, Al{\'u}, and Jankovic]{sakotic2023non}
Sakotic,~Z.; Stankovic,~P.; Bengin,~V.; Krasnok,~A.; Al{\'u},~A.; Jankovic,~N. Non-Hermitian Control of Topological Scattering Singularities Emerging from Bound States in the Continuum. \emph{Laser \& Photonics Reviews} \textbf{2023}, 2200308\relax
\mciteBstWouldAddEndPuncttrue
\mciteSetBstMidEndSepPunct{\mcitedefaultmidpunct}
{\mcitedefaultendpunct}{\mcitedefaultseppunct}\relax
\EndOfBibitem
\bibitem[Colom \latin{et~al.}(2023)Colom, Mikheeva, Achouri, Zuniga-Perez, Bonod, Martin, Burger, and Genevet]{colom2023crossing}
Colom,~R.; Mikheeva,~E.; Achouri,~K.; Zuniga-Perez,~J.; Bonod,~N.; Martin,~O. J.~F.; Burger,~S.; Genevet,~P. Crossing of the Branch Cut: The Topological Origin of a Universal 2$\pi$-Phase Retardation in Non-Hermitian Metasurfaces. \emph{Laser \& Photonics Reviews} \textbf{2023}, 2200976\relax
\mciteBstWouldAddEndPuncttrue
\mciteSetBstMidEndSepPunct{\mcitedefaultmidpunct}
{\mcitedefaultendpunct}{\mcitedefaultseppunct}\relax
\EndOfBibitem
\bibitem[Liu \latin{et~al.}(2023)Liu, Chen, Hu, Fan, Christodoulides, Zhao, and Qiu]{Liu2023TopologicalPA}
Liu,~M.; Chen,~W.; Hu,~G.; Fan,~S.; Christodoulides,~D.~N.; Zhao,~C.; Qiu,~C.-W. Spectral phase singularity and topological behavior in perfect absorption. \emph{Phys. Rev. B} \textbf{2023}, \emph{107}, L241403\relax
\mciteBstWouldAddEndPuncttrue
\mciteSetBstMidEndSepPunct{\mcitedefaultmidpunct}
{\mcitedefaultendpunct}{\mcitedefaultseppunct}\relax
\EndOfBibitem
\bibitem[Jariwala \latin{et~al.}(2016)Jariwala, Davoyan, Tagliabue, Sherrott, Wong, and Atwater]{jariwala2016near}
Jariwala,~D.; Davoyan,~A.~R.; Tagliabue,~G.; Sherrott,~M.~C.; Wong,~J.; Atwater,~H.~A. Near-unity absorption in van der Waals semiconductors for ultrathin optoelectronics. \emph{Nano letters} \textbf{2016}, \emph{16}, 5482--5487\relax
\mciteBstWouldAddEndPuncttrue
\mciteSetBstMidEndSepPunct{\mcitedefaultmidpunct}
{\mcitedefaultendpunct}{\mcitedefaultseppunct}\relax
\EndOfBibitem
\bibitem[Epstein \latin{et~al.}(2020)Epstein, Terr{\'e}s, Chaves, Pusapati, Rhodes, Frank, Zimmermann, Qin, Watanabe, Taniguchi, \latin{et~al.} others]{epstein2020near}
Epstein,~I.; Terr{\'e}s,~B.; Chaves,~A.~J.; Pusapati,~V.-V.; Rhodes,~D.~A.; Frank,~B.; Zimmermann,~V.; Qin,~Y.; Watanabe,~K.; Taniguchi,~T.; others Near-unity light absorption in a monolayer WS2 van der Waals heterostructure cavity. \emph{Nano letters} \textbf{2020}, \emph{20}, 3545--3552\relax
\mciteBstWouldAddEndPuncttrue
\mciteSetBstMidEndSepPunct{\mcitedefaultmidpunct}
{\mcitedefaultendpunct}{\mcitedefaultseppunct}\relax
\EndOfBibitem
\bibitem[Horng \latin{et~al.}(2020)Horng, Martin, Chou, Courtade, Chang, Hsu, Wentzel, Ruth, Lu, Cundiff, \latin{et~al.} others]{horng2020perfect}
Horng,~J.; Martin,~E.~W.; Chou,~Y.-H.; Courtade,~E.; Chang,~T.-c.; Hsu,~C.-Y.; Wentzel,~M.-H.; Ruth,~H.~G.; Lu,~T.-c.; Cundiff,~S.~T.; others Perfect absorption by an atomically thin crystal. \emph{Physical Review Applied} \textbf{2020}, \emph{14}, 024009\relax
\mciteBstWouldAddEndPuncttrue
\mciteSetBstMidEndSepPunct{\mcitedefaultmidpunct}
{\mcitedefaultendpunct}{\mcitedefaultseppunct}\relax
\EndOfBibitem
\bibitem[Ermolaev \latin{et~al.}(2022)Ermolaev, Voronin, Baranov, Kravets, Tselikov, Stebunov, Yakubovsky, Novikov, Vyshnevyy, Mazitov, \latin{et~al.} others]{ermolaev2022topological}
Ermolaev,~G.; Voronin,~K.; Baranov,~D.~G.; Kravets,~V.; Tselikov,~G.; Stebunov,~Y.; Yakubovsky,~D.; Novikov,~S.; Vyshnevyy,~A.; Mazitov,~A.; others Topological phase singularities in atomically thin high-refractive-index materials. \emph{Nature Communications} \textbf{2022}, \emph{13}, 2049\relax
\mciteBstWouldAddEndPuncttrue
\mciteSetBstMidEndSepPunct{\mcitedefaultmidpunct}
{\mcitedefaultendpunct}{\mcitedefaultseppunct}\relax
\EndOfBibitem
\bibitem[Canales \latin{et~al.}(2023)Canales, Kotov, and Shegai]{canales2023perfect}
Canales,~A.; Kotov,~O.; Shegai,~T.~O. Perfect Absorption and Strong Coupling in Supported MoS2 Multilayers. \emph{ACS nano} \textbf{2023}, \emph{17}, 3401--3411\relax
\mciteBstWouldAddEndPuncttrue
\mciteSetBstMidEndSepPunct{\mcitedefaultmidpunct}
{\mcitedefaultendpunct}{\mcitedefaultseppunct}\relax
\EndOfBibitem
\bibitem[Alcaraz~Iranzo \latin{et~al.}(2018)Alcaraz~Iranzo, Nanot, Dias, Epstein, Peng, Efetov, Lundeberg, Parret, Osmond, Hong, \latin{et~al.} others]{alcaraz2018probing}
Alcaraz~Iranzo,~D.; Nanot,~S.; Dias,~E.~J.; Epstein,~I.; Peng,~C.; Efetov,~D.~K.; Lundeberg,~M.~B.; Parret,~R.; Osmond,~J.; Hong,~J.-Y.; others Probing the ultimate plasmon confinement limits with a van der Waals heterostructure. \emph{Science} \textbf{2018}, \emph{360}, 291--295\relax
\mciteBstWouldAddEndPuncttrue
\mciteSetBstMidEndSepPunct{\mcitedefaultmidpunct}
{\mcitedefaultendpunct}{\mcitedefaultseppunct}\relax
\EndOfBibitem
\bibitem[Rivera \latin{et~al.}(2019)Rivera, Christensen, and Narang]{rivera2019phonon}
Rivera,~N.; Christensen,~T.; Narang,~P. Phonon polaritonics in two-dimensional materials. \emph{Nano Letters} \textbf{2019}, \emph{19}, 2653--2660\relax
\mciteBstWouldAddEndPuncttrue
\mciteSetBstMidEndSepPunct{\mcitedefaultmidpunct}
{\mcitedefaultendpunct}{\mcitedefaultseppunct}\relax
\EndOfBibitem
\bibitem[Ma \latin{et~al.}(2022)Ma, Hu, Waldecker, Watanabe, Taniguchi, Liu, and Heinz]{ma2022reststrahlen}
Ma,~E.~Y.; Hu,~J.; Waldecker,~L.; Watanabe,~K.; Taniguchi,~T.; Liu,~F.; Heinz,~T.~F. The Reststrahlen Effect in the Optically Thin Limit: A Framework for Resonant Response in Thin Media. \emph{Nano Letters} \textbf{2022}, \emph{22}, 8389--8393\relax
\mciteBstWouldAddEndPuncttrue
\mciteSetBstMidEndSepPunct{\mcitedefaultmidpunct}
{\mcitedefaultendpunct}{\mcitedefaultseppunct}\relax
\EndOfBibitem
\bibitem[Luhmann \latin{et~al.}(2020)Luhmann, H{\o}j, Piller, K{\"a}hler, Chien, West, Andersen, and Schmid]{luhmann2020ultrathin}
Luhmann,~N.; H{\o}j,~D.; Piller,~M.; K{\"a}hler,~H.; Chien,~M.-H.; West,~R.~G.; Andersen,~U.~L.; Schmid,~S. Ultrathin 2 nm gold as impedance-matched absorber for infrared light. \emph{Nature Communications} \textbf{2020}, \emph{11}, 2161\relax
\mciteBstWouldAddEndPuncttrue
\mciteSetBstMidEndSepPunct{\mcitedefaultmidpunct}
{\mcitedefaultendpunct}{\mcitedefaultseppunct}\relax
\EndOfBibitem
\bibitem[Zhao \latin{et~al.}(2023)Zhao, Xie, Wan, Ding, Liu, Xie, Li, Chen, Wang, Zhang, \latin{et~al.} others]{zhao2023ultrathin}
Zhao,~T.; Xie,~P.; Wan,~H.; Ding,~T.; Liu,~M.; Xie,~J.; Li,~E.; Chen,~X.; Wang,~T.; Zhang,~Q.; others Ultrathin MXene assemblies approach the intrinsic absorption limit in the 0.5--10 THz band. \emph{Nature Photonics} \textbf{2023}, 1--7\relax
\mciteBstWouldAddEndPuncttrue
\mciteSetBstMidEndSepPunct{\mcitedefaultmidpunct}
{\mcitedefaultendpunct}{\mcitedefaultseppunct}\relax
\EndOfBibitem
\bibitem[Huo and Konstantatos(2018)Huo, and Konstantatos]{huo2018recent}
Huo,~N.; Konstantatos,~G. Recent progress and future prospects of 2D-based photodetectors. \emph{Advanced Materials} \textbf{2018}, \emph{30}, 1801164\relax
\mciteBstWouldAddEndPuncttrue
\mciteSetBstMidEndSepPunct{\mcitedefaultmidpunct}
{\mcitedefaultendpunct}{\mcitedefaultseppunct}\relax
\EndOfBibitem
\bibitem[Akinwande \latin{et~al.}(2019)Akinwande, Huyghebaert, Wang, Serna, Goossens, Li, Wong, and Koppens]{akinwande2019graphene}
Akinwande,~D.; Huyghebaert,~C.; Wang,~C.-H.; Serna,~M.~I.; Goossens,~S.; Li,~L.-J.; Wong,~H.-S.~P.; Koppens,~F.~H. Graphene and two-dimensional materials for silicon technology. \emph{Nature} \textbf{2019}, \emph{573}, 507--518\relax
\mciteBstWouldAddEndPuncttrue
\mciteSetBstMidEndSepPunct{\mcitedefaultmidpunct}
{\mcitedefaultendpunct}{\mcitedefaultseppunct}\relax
\EndOfBibitem
\bibitem[Turunen \latin{et~al.}(2022)Turunen, Brotons-Gisbert, Dai, Wang, Scerri, Bonato, J{\"o}ns, Sun, and Gerardot]{turunen2022quantum}
Turunen,~M.; Brotons-Gisbert,~M.; Dai,~Y.; Wang,~Y.; Scerri,~E.; Bonato,~C.; J{\"o}ns,~K.~D.; Sun,~Z.; Gerardot,~B.~D. Quantum photonics with layered 2D materials. \emph{Nature Reviews Physics} \textbf{2022}, \emph{4}, 219--236\relax
\mciteBstWouldAddEndPuncttrue
\mciteSetBstMidEndSepPunct{\mcitedefaultmidpunct}
{\mcitedefaultendpunct}{\mcitedefaultseppunct}\relax
\EndOfBibitem
\bibitem[Yu \latin{et~al.}(2010)Yu, Raman, and Fan]{yu2010fundamental}
Yu,~Z.; Raman,~A.; Fan,~S. Fundamental limit of nanophotonic light trapping in solar cells. \emph{Proceedings of the National Academy of Sciences} \textbf{2010}, \emph{107}, 17491--17496\relax
\mciteBstWouldAddEndPuncttrue
\mciteSetBstMidEndSepPunct{\mcitedefaultmidpunct}
{\mcitedefaultendpunct}{\mcitedefaultseppunct}\relax
\EndOfBibitem
\bibitem[Miller \latin{et~al.}(2016)Miller, Polimeridis, Reid, Hsu, DeLacy, Joannopoulos, Solja{\v{c}}i{\'c}, and Johnson]{miller2016fundamental}
Miller,~O.~D.; Polimeridis,~A.~G.; Reid,~M.~H.; Hsu,~C.~W.; DeLacy,~B.~G.; Joannopoulos,~J.~D.; Solja{\v{c}}i{\'c},~M.; Johnson,~S.~G. Fundamental limits to optical response in absorptive systems. \emph{Optics express} \textbf{2016}, \emph{24}, 3329--3364\relax
\mciteBstWouldAddEndPuncttrue
\mciteSetBstMidEndSepPunct{\mcitedefaultmidpunct}
{\mcitedefaultendpunct}{\mcitedefaultseppunct}\relax
\EndOfBibitem
\bibitem[Molesky \latin{et~al.}(2019)Molesky, Jin, Venkataram, and Rodriguez]{molesky2019t}
Molesky,~S.; Jin,~W.; Venkataram,~P.~S.; Rodriguez,~A.~W. T operator bounds on angle-integrated absorption and thermal radiation for arbitrary objects. \emph{Physical review letters} \textbf{2019}, \emph{123}, 257401\relax
\mciteBstWouldAddEndPuncttrue
\mciteSetBstMidEndSepPunct{\mcitedefaultmidpunct}
{\mcitedefaultendpunct}{\mcitedefaultseppunct}\relax
\EndOfBibitem
\bibitem[Kuang \latin{et~al.}(2020)Kuang, Zhang, and Miller]{kuang2020maximal}
Kuang,~Z.; Zhang,~L.; Miller,~O.~D. Maximal single-frequency electromagnetic response. \emph{Optica} \textbf{2020}, \emph{7}, 1746--1757\relax
\mciteBstWouldAddEndPuncttrue
\mciteSetBstMidEndSepPunct{\mcitedefaultmidpunct}
{\mcitedefaultendpunct}{\mcitedefaultseppunct}\relax
\EndOfBibitem
\bibitem[Chao \latin{et~al.}(2022)Chao, Strekha, Kuate~Defo, Molesky, and Rodriguez]{chao2022physical}
Chao,~P.; Strekha,~B.; Kuate~Defo,~R.; Molesky,~S.; Rodriguez,~A.~W. Physical limits in electromagnetism. \emph{Nature Reviews Physics} \textbf{2022}, \emph{4}, 543--559\relax
\mciteBstWouldAddEndPuncttrue
\mciteSetBstMidEndSepPunct{\mcitedefaultmidpunct}
{\mcitedefaultendpunct}{\mcitedefaultseppunct}\relax
\EndOfBibitem
\bibitem[Li and Heinz(2018)Li, and Heinz]{li2018two}
Li,~Y.; Heinz,~T.~F. Two-dimensional models for the optical response of thin films. \emph{2D Materials} \textbf{2018}, \emph{5}, 025021\relax
\mciteBstWouldAddEndPuncttrue
\mciteSetBstMidEndSepPunct{\mcitedefaultmidpunct}
{\mcitedefaultendpunct}{\mcitedefaultseppunct}\relax
\EndOfBibitem
\bibitem[Pu \latin{et~al.}(2012)Pu, Feng, Wang, Hu, Huang, Ma, Zhao, Wang, and Luo]{pu2012ultrathin}
Pu,~M.; Feng,~Q.; Wang,~M.; Hu,~C.; Huang,~C.; Ma,~X.; Zhao,~Z.; Wang,~C.; Luo,~X. Ultrathin broadband nearly perfect absorber with symmetrical coherent illumination. \emph{Optics express} \textbf{2012}, \emph{20}, 2246--2254\relax
\mciteBstWouldAddEndPuncttrue
\mciteSetBstMidEndSepPunct{\mcitedefaultmidpunct}
{\mcitedefaultendpunct}{\mcitedefaultseppunct}\relax
\EndOfBibitem
\bibitem[Luo \latin{et~al.}(2014)Luo, Li, Hou, and Lai]{luo2014unified}
Luo,~J.; Li,~S.; Hou,~B.; Lai,~Y. Unified theory for perfect absorption in ultrathin absorptive films with constant tangential electric or magnetic fields. \emph{Physical Review B} \textbf{2014}, \emph{90}, 165128\relax
\mciteBstWouldAddEndPuncttrue
\mciteSetBstMidEndSepPunct{\mcitedefaultmidpunct}
{\mcitedefaultendpunct}{\mcitedefaultseppunct}\relax
\EndOfBibitem
\bibitem[Li \latin{et~al.}(2014)Li, Luo, Anwar, Li, Lu, Hang, Lai, Hou, Shen, and Wang]{li2014equivalent}
Li,~S.; Luo,~J.; Anwar,~S.; Li,~S.; Lu,~W.; Hang,~Z.~H.; Lai,~Y.; Hou,~B.; Shen,~M.; Wang,~C. An equivalent realization of coherent perfect absorption under single beam illumination. \emph{Scientific reports} \textbf{2014}, \emph{4}, 7369\relax
\mciteBstWouldAddEndPuncttrue
\mciteSetBstMidEndSepPunct{\mcitedefaultmidpunct}
{\mcitedefaultendpunct}{\mcitedefaultseppunct}\relax
\EndOfBibitem
\bibitem[Harbecke \latin{et~al.}(1985)Harbecke, Heinz, and Grosse]{harbecke1985optical}
Harbecke,~B.; Heinz,~B.; Grosse,~P. Optical properties of thin films and the Berreman effect. \emph{Applied Physics A} \textbf{1985}, \emph{38}, 263--267\relax
\mciteBstWouldAddEndPuncttrue
\mciteSetBstMidEndSepPunct{\mcitedefaultmidpunct}
{\mcitedefaultendpunct}{\mcitedefaultseppunct}\relax
\EndOfBibitem
\bibitem[Luo \latin{et~al.}(2018)Luo, Tsai, Gu, and Hong]{luo2018subwavelength}
Luo,~X.; Tsai,~D.; Gu,~M.; Hong,~M. Subwavelength interference of light on structured surfaces. \emph{Advances in Optics and Photonics} \textbf{2018}, \emph{10}, 757--842\relax
\mciteBstWouldAddEndPuncttrue
\mciteSetBstMidEndSepPunct{\mcitedefaultmidpunct}
{\mcitedefaultendpunct}{\mcitedefaultseppunct}\relax
\EndOfBibitem
\bibitem[Kim \latin{et~al.}(2016)Kim, Park, Esfandyarpour, Pecora, Kik, and Brongersma]{kim2016superabsorbing}
Kim,~S.~J.; Park,~J.; Esfandyarpour,~M.; Pecora,~E.~F.; Kik,~P.~G.; Brongersma,~M.~L. Superabsorbing, artificial metal films constructed from semiconductor nanoantennas. \emph{Nano Letters} \textbf{2016}, \emph{16}, 3801--3808\relax
\mciteBstWouldAddEndPuncttrue
\mciteSetBstMidEndSepPunct{\mcitedefaultmidpunct}
{\mcitedefaultendpunct}{\mcitedefaultseppunct}\relax
\EndOfBibitem
\bibitem[Beliaev \latin{et~al.}(2021)Beliaev, Shkondin, Lavrinenko, and Takayama]{beliaev2021thickness}
Beliaev,~L.~Y.; Shkondin,~E.; Lavrinenko,~A.~V.; Takayama,~O. Thickness-dependent optical properties of aluminum nitride films for mid-infrared wavelengths. \emph{Journal of Vacuum Science \& Technology A: Vacuum, Surfaces, and Films} \textbf{2021}, \emph{39}, 043408\relax
\mciteBstWouldAddEndPuncttrue
\mciteSetBstMidEndSepPunct{\mcitedefaultmidpunct}
{\mcitedefaultendpunct}{\mcitedefaultseppunct}\relax
\EndOfBibitem
\bibitem[Siegman(1986)]{siegman1986lasers}
Siegman,~A.~E. \emph{Lasers}; University science books, 1986\relax
\mciteBstWouldAddEndPuncttrue
\mciteSetBstMidEndSepPunct{\mcitedefaultmidpunct}
{\mcitedefaultendpunct}{\mcitedefaultseppunct}\relax
\EndOfBibitem
\bibitem[Rozanov(2000)]{rozanov2000ultimate}
Rozanov,~K.~N. Ultimate thickness to bandwidth ratio of radar absorbers. \emph{IEEE Transactions on Antennas and Propagation} \textbf{2000}, \emph{48}, 1230--1234\relax
\mciteBstWouldAddEndPuncttrue
\mciteSetBstMidEndSepPunct{\mcitedefaultmidpunct}
{\mcitedefaultendpunct}{\mcitedefaultseppunct}\relax
\EndOfBibitem
\bibitem[Venkataram \latin{et~al.}(2020)Venkataram, Molesky, Jin, and Rodriguez]{venkataram2020fundamental}
Venkataram,~P.~S.; Molesky,~S.; Jin,~W.; Rodriguez,~A.~W. Fundamental limits to radiative heat transfer: The limited role of nanostructuring in the near-field. \emph{Physical review letters} \textbf{2020}, \emph{124}, 013904\relax
\mciteBstWouldAddEndPuncttrue
\mciteSetBstMidEndSepPunct{\mcitedefaultmidpunct}
{\mcitedefaultendpunct}{\mcitedefaultseppunct}\relax
\EndOfBibitem
\end{mcitethebibliography}

\end{document}